\documentclass[%
 aip,
 amsmath,amssymb,
 reprint,%
]{revtex4-1}

\usepackage{graphicx}% Include figure files
\usepackage{dcolumn}% Align table columns on decimal point
\usepackage{bm}% bold math
\usepackage[utf8]{inputenc}
\usepackage[T1]{fontenc}
\usepackage{mathptmx}
\usepackage{etoolbox}

\makeatletter
\def\@email#1#2{%
 \endgroup
 \patchcmd{\titleblock@produce}
  {\frontmatter@RRAPformat}
  {\frontmatter@RRAPformat{\produce@RRAP{*#1\href{mailto:#2}{#2}}}\frontmatter@RRAPformat}
  {}{}
}%
\makeatother
\begin{document}

\preprint{AIP/123-QED}

\title[Instability of the discontinuity]{Two-dimensional particle simulation of the boundary between a hot pair plasma and magnetized electrons and protons: out-of-plane magnetic field}
% Force line breaks with \\
\author{M.~E.~Dieckmann}
\affiliation{Department of Science and Technology (ITN), Link\"oping University, 60174 Norrk\"oping, Sweden}
\email{mark.e.dieckmann@liu.se}
\author{D.~Folini}
\author{R.~Walder}
\affiliation{Univ Lyon, ENS de Lyon, Univ Lyon 1,  CNRS, Centre de Recherche Astrophysique de Lyon UMR5574, F-69230, Saint-Genis-Laval, France}
\author{A.~Charlet}
\affiliation{Univ Lyon, ENS de Lyon, Univ Lyon 1,  CNRS, Centre de Recherche Astrophysique de Lyon UMR5574, F-69230, Saint-Genis-Laval, France}
\affiliation{Laboratoire Univers et Particules de Montpellier (LUPM), Universit\'e de Montpellier, CNRS/IN2P3, CC72, place Eug\`ene Bataillon, F-34095 Montpellier Cedex 5, France}
\affiliation{Astrophysics Research Center of the Open University (ARCO), The Open University of Israel, P.O. Box 808, Ra’anana 4353701, Israel}
\author{A.~Marcowith}%
\affiliation{Laboratoire Univers et Particules de Montpellier (LUPM), Universit\'e de Montpellier, CNRS/IN2P3, CC72, place Eug\`ene Bataillon, F-34095 Montpellier Cedex 5, France}

\date{\today}% It is always \today, today,

\begin{abstract}
By means of a particle-in-cell (PIC) simulation, we study the interaction between a uniform magnetized ambient electron-proton plasma at rest and an unmagnetized pair plasma, which we inject at one simulation boundary with a mildly relativistic mean speed and temperature. The magnetic field points out of the simulation plane. The injected pair plasma expels the magnetic field and piles it up at its front. It traps ambient electrons and drags them across the protons. An electric field grows, which accelerates protons into the pair cloud's expansion direction. This electromagnetic pulse separates the pair cloud from the ambient plasma. Electrons and positrons, which drift in the pulse's nonuniform field, trigger an instability that disrupts the current sheet ahead of the pulse. The wave vector of the growing perturbation is orthogonal to the magnetic field direction and magnetic tension cannot stabilize it. The electromagnetic pulse becomes permeable for pair plasma, which forms new electromagnetic pulses ahead of the initial one. A transition layer develops with a thickness of a few proton skin depths, in which protons and positrons are accelerated by strong electromagnetic fields. Protons form dense clumps surrounded by a strong magnetic field. The thickness of the transition layer grows less rapidly than we would expect from the typical speeds of the pair plasma particles and the latter transfer momentum to protons; hence, the transition layer acts as a discontinuity, separating the pair plasma from the ambient plasma. Such a discontinuity is an important building block for astrophysical pair plasma jets.
\end{abstract}
\maketitle

\section{\label{intro}Introduction}

Some binary systems consisting of an accreting neutron star or black hole and a companion star are sources of pair plasma.~\cite{Mirabel94,Siegert16} The pair plasma is generated by energetic processes near the inner accretion disc or through the interaction of electromagnetic radiation with the strong intrinsic magnetic fields of the compact object or its accretion disc.~\cite{Blandford77,Blandford82,Yuan14} The ejected pair plasma must eventually interact with the wind of the stellar companion. This interaction can channel the pair outflow into a jet that can reach a superluminal speed.\cite{Mirabel94,Fender04} Such pair outflows have been named as a possible source for galactic positrons.\cite{Prantzos11} Hydrodynamic models provide an intuitive description of the jet structure.~\cite{Aloy99,Bromberg11,Perucho08,Charlet22} They apply if the mean free path of the particles, which constitute the fluid, is small compared to the spatial scales of interest. Hydrodynamic models take into account important elementary structures like sound- and rarefaction waves, shocks, and contact discontinuities. Contact discontinuities separate two fluids of different origin, composition, density, and temperature.  

\begin{figure}[ht]
\includegraphics[width=\columnwidth]{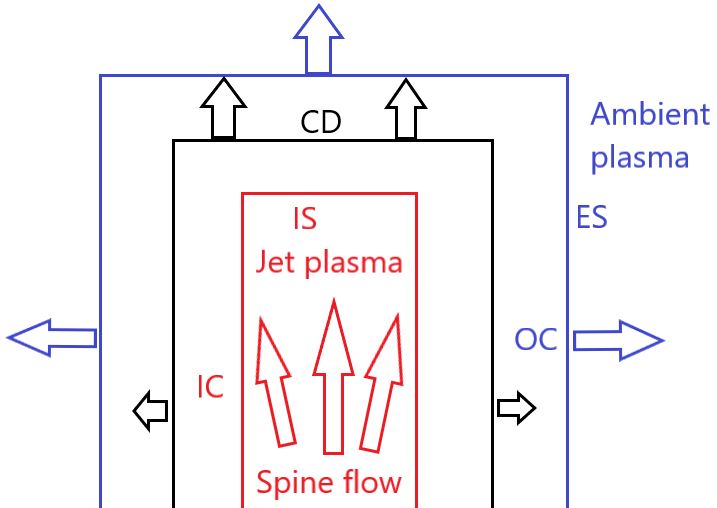}
\caption{Schematic jet structure: A contact discontinuity (CD) separates the jet's pair plasma from the ambient electron-proton plasma. The spine flow has a low mass density and a large speed. It is slowed down, heated, and compressed as it crosses the internal shock (IS) and flows into the inner cocoon (IC) between the IS and the CD. As indicated by the arrows, the hot plasma in the IC imposes pressure onto the CD through which it expands outward. It pushes the ambient plasma ahead of it. An outward-propagating external shock (ES) separates the ambient plasma from the moving ambient plasma in the outer cocoon (OC) between the CD and ES.}
\label{figure1}
\end{figure}

Figure~\ref{figure1} is a sketch of a hydrodynamic jet near its front. The plasma in the spine flow has a low mass density and a high bulk velocity. The ambient plasma far from the jet is at rest. A contact discontinuity (CD) separates both fluids. In what follows, we assume that all material enclosed by the CD is pair plasma while all material outside the CD is plasma composed of protons and electrons. We can thus distinguish both fluids by the carrier of positive charge. The streaming pair plasma is slowed down, compressed, and heated up as it approaches the CD forming a layer of hot material near it; the inner cocoon. An internal shock develops between the inner cocoon and the spine flow if the pair plasma's mean speed change exceeds the local sound speed. The thermal pressure of the hot plasma in the inner cocoon pushes the CD away from the spine flow. The moving CD accelerates the nearby ambient plasma. A shock forms at the front of the accelerated ambient material if its speed exceeds the local sound speed. Most of the jet's momentum is transferred to the CD at the jet's head, giving the jet its characteristic elongated shape.  

Hydrodynamic models assume that CDs and shocks are thin compared to the scales of interest. The mean free path of particles in the interstellar medium or a stellar wind is, however, not negligibly small compared to the jet size. Therefore, binary collisions are replaced by mechanisms based on the electromagnetic fields induced by the collective motion of the plasma particles, as the means to exchange momentum and energy between particles. Certain properties of shocks and discontinuities in such material, which is known as collisionless plasma, are different from those of their hydrodynamic counterparts with potentially far-reaching astrophysical consequences. It is important to determine if and how electromagnetic fields can sustain discontinuities in a collisionless plasma and if the discontinuities remain thin compared to the spatial scales of interest. 

Particle-in-cell (PIC) simulations can resolve all structures in collisionless plasma. Most previous PIC simulation studies related to jets in collisionless plasma have focused on pair plasma shocks\cite{Nishikawa03,Chang08} and electron-ion shocks,\cite{Frederiksen04,Spitkovsky08a} which correspond to the internal and external shocks in Fig.~\ref{figure1}. One finding was that magnetic fields in the transition layers of relativistic shocks can have energy densities that exceed those expected from compression of the upstream magnetic field. This magnetic energy can be released through magnetic reconnection.~\cite{Melzani14,Marcowith16} Another result is that collisionless shocks can accelerate a small fraction of plasma particles to cosmic ray energies.~\cite{Marcowith16} The onset of such an acceleration has been studied with PIC simulations.~\cite{Spitkovsky08b} Discontinuities between pair plasma and electron-proton plasma like the CD in Fig.~\ref{figure1} have not been explored to the same extent. Such a discontinuity has been observed in two-dimensional PIC simulations of a pair plasma that propagated through a magnetized electron-proton plasma~\cite{Dieckmann19,Dieckmannetal20} and studied in one spatial dimension~\cite{Dieckmann20} assuming that it is planar. In the two-dimensional simulation,~\cite{Dieckmannetal20} the discontinuity became unstable to a magnetic Rayleigh-Taylor-like instability.~\cite{Winske96,Stone07,Bret11,Liu19,Hillier16} The wave vector of the perturbation was parallel to the magnetic field and the instability increased magnetic tension, which eventually quenched the instability.  

Here we examine the interface between an expanding unmagnetized pair plasma and a magnetized electron-proton plasma at rest using a two-dimensional particle-in-cell (PIC) simulation. We let the interface, which takes the role of the CD in Fig.~\ref{figure1}, grow self-consistently. The ambient electron-proton plasma is permeated by a spatially uniform magnetic field, which is oriented perpendicularly to the expansion direction of the pair cloud. Its magnetic pressure matches the electron thermal pressure. We use the same plasma conditions as in a previous simulation\cite{Dieckmannetal20} apart from a lower mean speed of the pair cloud and a magnetic field direction, which is now normal to the simulation plane. We obtain the following results. The expanding pair plasma expels the magnetic field and piles it up at its front. The moving magnetic field traps the electrons of the ambient plasma and pushes them into the pair plasma's expansion direction. Their current induces an electric field, which accelerates the protons to a speed comparable to the interface's speed. In what follows we refer to this interface as the electromagnetic pulse (EMP). It separates positrons from protons and resembles the one observed previously~\cite{Dieckmannetal20} at early simulation times. 

The magnetic field points out of the simulation plane and can be rearranged by a perturbation without bending field lines. Interchange modes of the magnetic Rayleigh-Taylor instability, which grow in such a plasma configuration, do not increase magnetic tension. Hence, they tend to be more unstable and disruptive than the undular mode~\cite{Liu19} studied in Ref.~\cite{Dieckmannetal20} We do not observe these interchange modes, because the EMP is destroyed before the magnetic Rayleigh-Taylor instability can set in. The current sheet, which confines the magnetic field ahead of the EMP, is sustained by ambient electrons that drift in the EMP's field and positrons that leaked through it. This current sheet is disrupted by a streaming instability between these particles and protons. The growing waves have a wavevector parallel to the drift direction of the ambient electrons, which was not resolved in the previous simulations.~\cite{Dieckmann19,Dieckmann20,Dieckmannetal20} These waves grow fast and their saturation lets the EMP become permeable for pair plasma, broadening the transition layer between the pair cloud and the ambient plasma. The speed, at which the thickness of the transition layer between the ambient plasma and the pair plasma increases, is well below the average speed of the pair plasma particles. This, together with the observed momentum transfer from the pair plasma to the ambient plasma, implies that the transition layer still acts as a discontinuity that is thin compared to the typical particle mean free paths near relativistic jets. 

The structure of our paper is as follows. Section 2 discusses the initial conditions of our simulation. Section 3 presents the early phase of the plasma collision while the late evolution is discussed in Section 4. Section 5 summarizes our findings.

\section{\label{setup}Initial conditions for the simulation}

Each plasma species in a collisionless plasma is represented by a phase space density distribution, which is a function of independent position and velocity coordinates. A PIC simulation code approximates the phase space fluid by an ensemble of computational particles (CPs), which have position and velocity coordinates and the same charge-to-mass ratio as the plasma species they represent. The current contributions of all CPs are summed up and give the macroscopic current density. This current density, the electric field, and the magnetic field are defined on a numerical grid and are connected via discretized forms of Maxwell's equations. The electromagnetic forces are interpolated to the position of each CP and update its velocity. We specify the time $T_{sim}$, during which we want to evolve the plasma, and the code subdivides it into time steps $\Delta_t$ with a duration that depends on the code's numerical scheme. The numerical cycle of the PIC code EPOCH we use is discussed in detail elsewhere.\cite{Arber2015} 

The two-dimensional simulation box is filled with an ambient plasma with the electron density $n_0$ and equally dense protons with the proton-to-electron mass ratio $m_p/m_e=$ 1836. Both species have the temperature $T_0=$ 2 keV. We normalize time to $\omega_{pi}^{-1}$ with the proton plasma frequency $\omega_{pi} = {(e^2n_0/\epsilon_0m_p)}^{1/2}$ ($e,c,\epsilon_0,\mu_0$: elementary charge, speed of light, vacuum permittivity and permeability). Space is normalized to the proton skin depth $\lambda_{i} = c/\omega_{pi}$. The simulation box with periodic boundaries resolves the spatial interval $L_x = 35$ along $x$ by 12000 grid cells and $L_y=8.75$ along $y$ by 3000 grid cells. A magnetic field with the amplitude $B_0$ is aligned with $z$. The electron thermal pressure $P_e = n_0k_B T_0$ ($k_B$: Boltzmann constant) equals the magnetic pressure $P_B=B_0^2/2\mu_0$. The proton gyrofrequency $\omega_{ci}=eB_0/(m_p\omega_{pi})$ is $2.1\times 10^{-3}$. We inject at the boundary $x=0$ a pair plasma, which consists of electrons and positrons with the temperature $50T_0$ (100 keV), has the mean speed $v_d=0.6c$ along increasing $x$, and the respective densities $n_0$ measured in the rest frame of the simulation box. 

We initialize the electrons and protons of the ambient plasma by 25 computational particles (CPs) per cell each. We inject 16 CPs per cell per time step to represent the electrons of the pair cloud and use the same number for the positrons. The simulation evolves the plasma until the final time $T_{sim}=200$ equivalent to 200$\omega_{pi}^{-1}$. 

In what follows, we present the data on a grid that is shifted relative to the simulation grid. Data from the simulation interval $L_x/2 \le x < L_x$ is moved to the the x-interval between $-L_x/2 \le x <0$. The boundary $x=0$, where we inject the pair cloud, is centered in the data grid. All displayed field components and plasma densities have been averaged over patches of 4 by 4 grid cells to improve the signal-to-noise ratio.

\section{\label{early}Initial evolution}

\begin{figure*}[ht]
    \includegraphics[width=\textwidth]{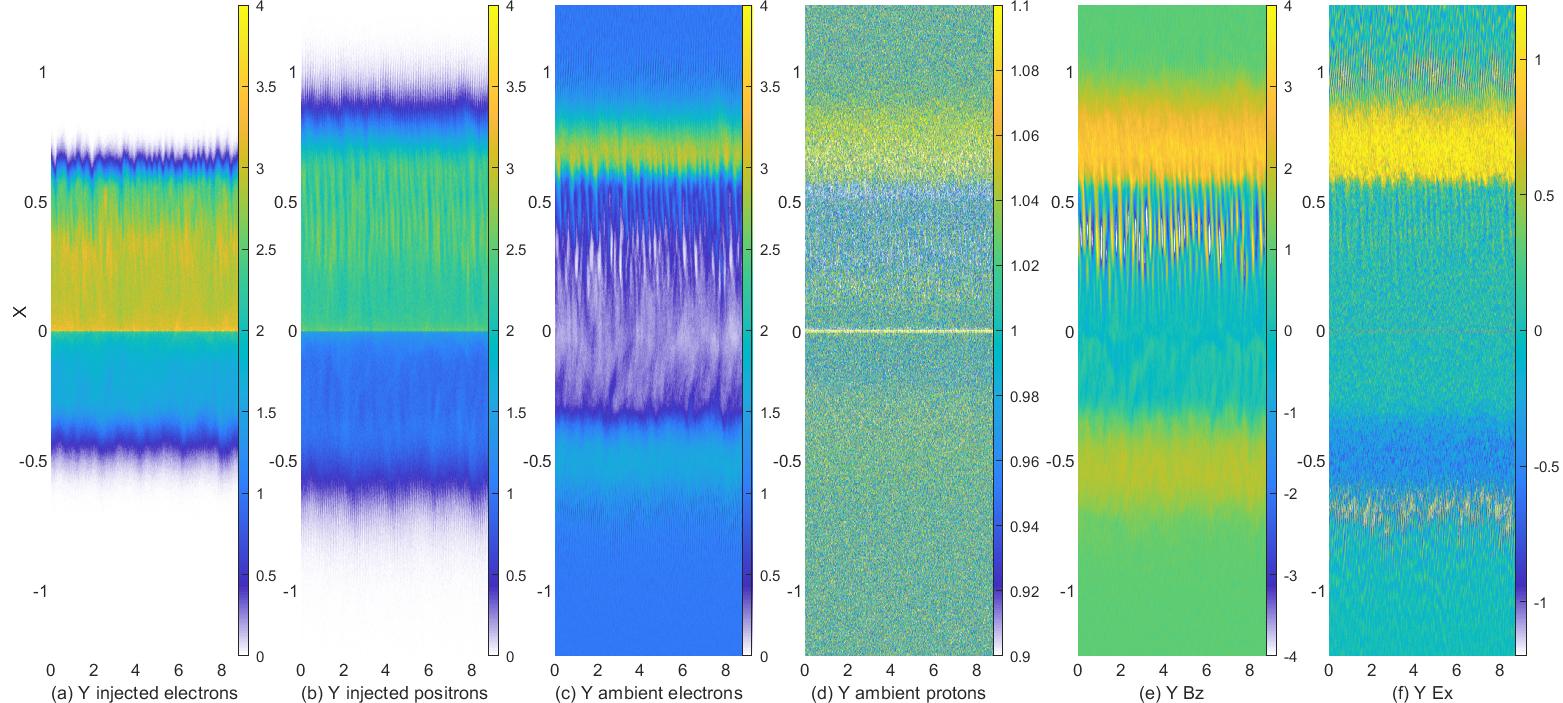}
    \caption{Plasma and field distribution at $t=5$: Panels (a, b) show the densities of the injected electrons and positrons. Densities of the ambient electrons and protons are displayed in panels (c, d). Panel (e) and (f) show $B_z$ and $E_x$. Densities are normalized to $n_0$, the magnetic field to $B_0$ and the electric field to $cB_0$.}
    \label{figure2}
\end{figure*}

Figure~\ref{figure2} shows the plasma and field distribution at the time $t=5$. 
The front of the injected electrons is located at $x \approx 0.6$ and that of the positrons at $x\approx 0.8$. Injected electrons occupy a smaller interval along $x$ than positrons and their density is higher. Most ambient electrons were expelled by the injected electrons and accumulated in the interval $0.6 \le x \le 0.8$. Pair cloud particles are scattered and reflected by the magnetized ambient plasma. Some return to the periodic boundary and cross it. Electrons in Fig.~\ref{figure2}(a) have expanded to $x\approx -0.45$ and positrons in Fig.~\ref{figure2}(b) to $x=-0.6$. Like in the case of the upward moving pair cloud front, the injected electrons are denser than the positrons and the ambient electrons accumulate ahead of them. The pair cloud, which expands in both directions from the injection boundary, is not symmetric around $x=0$. The injected pair cloud loses energy to the ambient plasma on its way up and the reflected particles interact with newly injected pair cloud particles as they return to the boundary. Hence, the pair cloud in the half-space $x<0$ starts its expansion later, it is closer to thermal equilibrium and it has a lower mean speed than the one in $x>0$.  

Positrons fill an interval in Fig.~\ref{figure2}(b) that is 1.5 wide. A particle with the mean speed $v_d$ of the pair cloud should have propagated the distance $5v_d/\lambda_i=3$ at the time shown in Fig.~\ref{figure2}. The slowdown of the pair cloud by the ambient plasma increases its density beyond 1. Figure~\ref{figure2}(d) reveals a pile-up of protons in the interval $0.6 \le x \le 0.9$, which is trailed by a depletion at lower $x$. Another barely visible proton accumulation is located in the interval $-0.6 \le x \le -0.25$. Figure~\ref{figure2}(e) reveals why injected and ambient electrons remain separated. The expanding pair cloud expels the magnetic field and piles it up at its front forming the structure we call EMP. The magnetic field is amplified to about 3$B_0$ in the interval $0.6 \le x \le 0.9$. The normalized gyroradius $r_{g}(v_0) = v_0m_e/(3eB_0\lambda_i)$ of an ambient electron in the amplified magnetic field with a speed $v_0$, which equals the thermal speed corresponding to the temperature 2 keV, is approximately $5.5 \times 10^{-3}$. This gyroradius is well below the thickness $\approx 0.3$ of the EMP. The gyroradius of leptons with $v_0=v_d$ is about $0.07$. The EMP is strong and wide enough to confine the injected pair cloud and trap the ambient electrons magnetically. 

\begin{figure}[ht]
\includegraphics[width=\columnwidth]{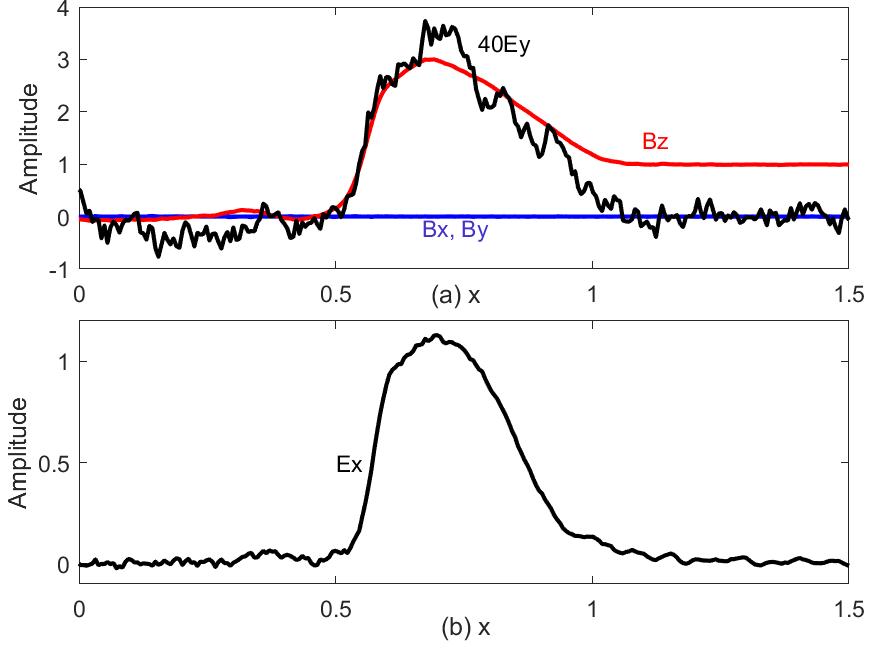}
\caption{Electromagnetic fields averaged over $y$ at the time $t=5$: We plot all magnetic field components and $40E_y$ in (a). We plot $E_x$ in~(b). Magnetic and electric fields are normalized to $B_0$ and $cB_0$.}
\label{figure3}
\end{figure}

Protons will only react to the electric field in Fig.~\ref{figure2}(f) since $\omega_{ci}T_{sim}=0.42$. We can estimate the speed to which they are accelerated once we know the EMP's propagation speed, which we determine with the help of the convective electric field. The magnetic field points along $z$ and the EMP propagates with the speed $v_p$ along $x$, which gives the convective electric field $E_y = v_p B_z$. We average the electric and magnetic field components over $y$ and plot them in Fig.~\ref{figure3}.
The magnetic $B_x$ and $B_y$ components oscillate around zero. The average $B_z$ vanishes for $0 \le x \le 0.5$ and equals $B_0$ for $x>1.1$. The y-averaged scaled $E_y$ component follows closely the EMP up to $x\approx 0.7$; its rear end moves at the speed $v_p\approx c/40$. The propagating EMP drags with it the ambient electrons. Their current drives the electric field in Fig.~\ref{figure2}(f). We estimate the proton velocity change $\Delta_v$ as follows. The average electric field $E_x \approx 1$ in Fig.~\ref{figure3}(b) corresponds to $E_{p,x}=c B_0$ in physical units. Protons are at rest before the EMP with the width $\Delta_p \approx 0.3\lambda_i$ and speed $v_p$ arrives. Their approximate exposure time to its electric field in physical units is $\delta_t = \Delta_p / v_p$ or $12 / \omega_{pi}$. The Lorentz force equation gives us the approximate velocity change in physical units 
\begin{equation}
\Delta_v \approx \frac{eE_{p,x}}{m_p} \delta_t = \frac{e c B_0}{m_p} \delta_t = 12 c\frac{\omega_{ci}}{\omega_{pi}}\approx v_p.
\label{estimate}
\end{equation}

\begin{figure*}[ht]
    \includegraphics[width=\textwidth]{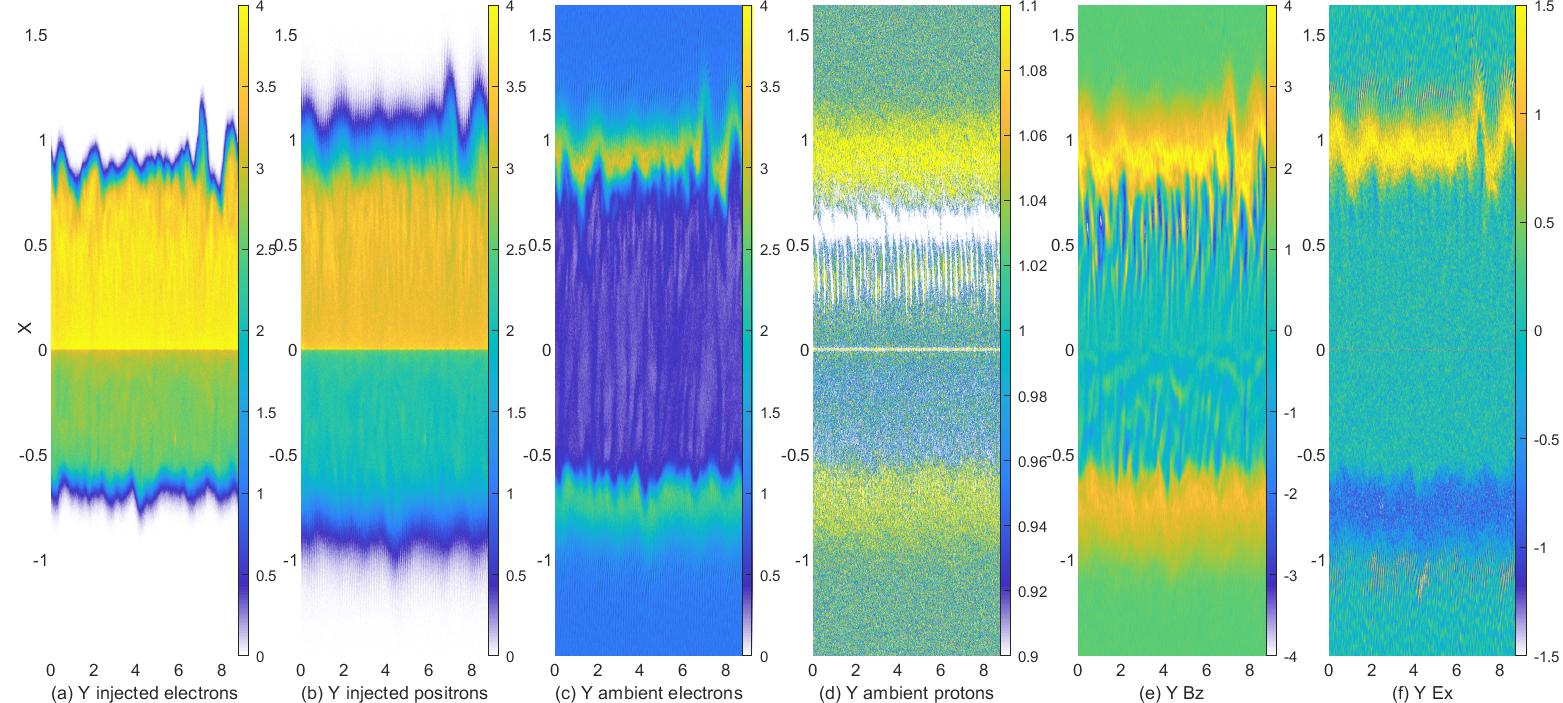}
    \caption{Plasma and field distribution at $t=10$: Panels (a, b) show the densities of the injected electrons and positrons. Densities of the ambient electrons and protons are displayed in panels (c, d). Panel (e) and (f) show $B_z$ and $E_x$. Densities are normalized to $n_0$, the magnetic field to $B_0$ and the electric field to $cB_0$.}
    \label{figure4}
\end{figure*}

Figure~\ref{figure4} considers the time $t=10$. The pair cloud is confined by EMPs on both sides of the boundary $x=0$. The central position along $x$ of each EMP oscillates as a function of $y$. The electric field points orthogonally to the front of the EMP, which is becoming increasingly distorted. It can thus also have a component $E_y\neq 0$. Most plots only show $E_x$ because we are primarily interested in how protons are accelerated along the mean expansion direction of the pair cloud.    
Figure~\ref{figure4}(d) evidences that the amplitude of the proton density modulation along $x$ has increased; the one near the EMP in the half-space $x>0$ has reached the amplitude $0.1$. The proton density modulation continues to increase. 

\begin{figure}[ht]
\includegraphics[width=\columnwidth]{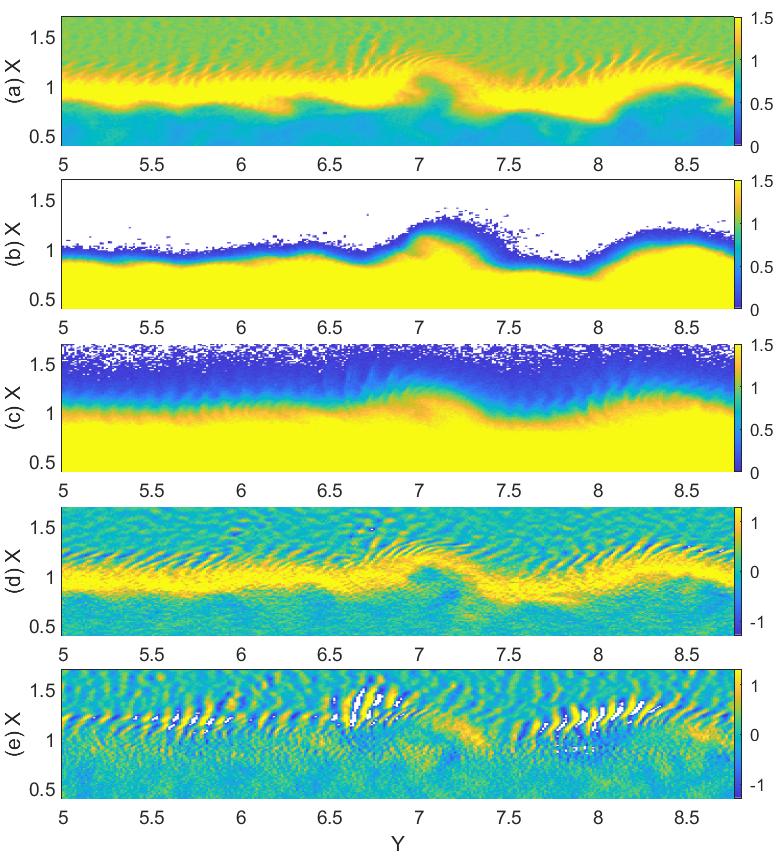}
\caption{The perturbed boundary at $t=10$: The square root of the normalized density of the ambient electrons is shown in (a). Those of the electrons and positrons of the pair cloud are shown in (b) and (c), respectively. Panels (d) and (e) show $E_x/cB_0$ and $E_y/cB_0$, respectively. Color scales in (a-c)  are clamped to the maximum value 1.5 and those in (d, e) to -1.3 and 1.3.}
\label{figure5}
\end{figure}

Figure~\ref{figure5} sheds light on the mechanism, that deformed the initial EMP in Fig.~\ref{figure4} and will eventually lead to its destruction. We focus on the distributions of the electrons, positrons, and relevant electric field components near the right part of the initial EMP in Fig.~\ref{figure4}. We display the square root of the lepton densities in Fig. ~\ref{figure5}(a-c). Ambient electrons in Fig.~\ref{figure5}(a) were piled up at $x\approx 1$ by the expanding pair cloud. 
The magnetic field of the EMP is strong enough to confine the bulk of the pair cloud. The sizeable force the $E_x$ component exerts on particles slows down the electrons of the pair cloud and accelerates its positrons. As a result, the electron density in Fig.~\ref{figure5}(b) decreases sharply at the transition layer and hardly any electrons reach $x>1.2$. Some positrons cross the EMP, and rotate in the homogeneous magnetic field of the ambient plasma until the moving EMP catches up with them. The motion of these particles gives rise to a net current along the negative $y$-direction ahead of the EMP. This current, together with that of the ambient electrons that drift in the electromagnetic field of the EMP and its gradient,  gives rise to the sharp decrease of the magnetic field amplitude at the EMP's front. 

The temperature of the ambient electrons and their mean speed $v_p \approx c/40$ relative to the EMP are well below the temperature and mean speed $v_d$ of the pair cloud; their gyroradius is thus much smaller than the width of the EMP and we can approximate their trajectory as gyrations around a drifting guiding center. Protons, on the other hand, are practically unmagnetized. We estimate the drift speed $v_{eb} = E_x/B_z$ of ambient electrons using the values $E_x \approx 1$ and $B_z \approx 2.5$ obtained from Fig.~\ref{figure4}. The drift speed $v_{eb}=0.4c$ is about 6 times larger than the initial thermal speed of the ambient electrons. The current due to the drifting electrons has the same direction as the contribution from the gyrating positrons and both add up. The mildly relativistic drift speed of the ambient electrons together with their large density ahead of the EM implies that they contribute most of the current ahead of the EMP. 

Such a fast relative drift between ions and magnetized ambient electrons leads to the growth of electrostatic upper-hybrid waves and electron-cyclotron waves.~\cite{Dieckmann00} They cause the electron-scale oscillations of the electric $E_x$ and $E_y$ components ahead of the EMP in Fig.~\ref{figure5}(d, e). These waves are strong enough to bunch positrons, as can be seen in Fig.~\ref{figure5}(c). This implies that the current density ahead of the EMP and, hence, the magnetic field gradient change with $y$.  Figure~\ref{figure5} demonstrates that the wave fields are not distributed uniformly along the front of the EMP and that their field amplitude exceeds that of the EMP. These waves are strong and nonuniform and electrostatic drift instabilities are thus the most likely reason for the EMP's deformation. Figure~\ref{figure5}(e) also shows electric field patches, like the one at $y\approx 7.25$ and $x\approx 1$, which are a consequence of the tilt of the EMP.

\begin{figure*}[ht]
    \includegraphics[width=\textwidth]{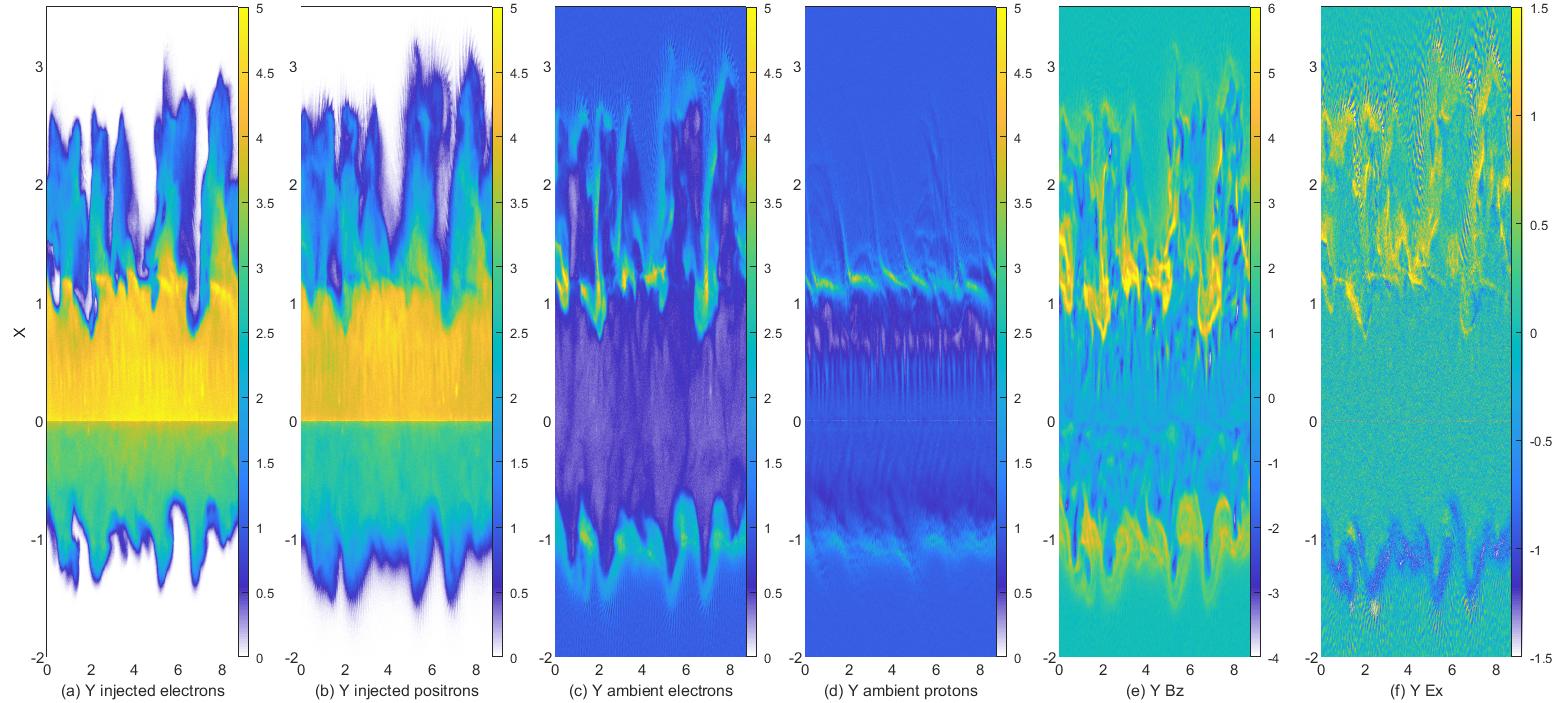}
    \caption{Plasma and field distribution at $t=20$: Panels (a, b) show the densities of the injected electrons and positrons. Densities of the ambient electrons and protons are displayed in panels (c, d). Panel (e) and (f) show $B_z$ and $E_x$. Densities are normalized to $n_0$, the magnetic field to $B_0$ and the electric field to $cB_0$.}
    \label{figure6}
\end{figure*}

The modulation of the EMP continues to grow and Fig.~\ref{figure6} reveals the consequences of this deformation. Fingers in the pair cloud have extended far beyond the EMP. The injected electrons have expelled ambient electrons and the pair cloud has piled up the magnetic field at the border of the fingers. The electric field, driven by the current of the expelled ambient electrons, has started to accelerate protons well upstream of the initial EMP. In what follows, we refer with initial EMP to the one that formed first. The term EMP refers to the electromagnetic pulse that marks the border between the pair cloud and unperturbed ambient plasma.
The reduced proton density behind the initial EMPs in Fig.~\ref{figure6}(d) proves their ability to accelerate protons. The initial EMP in the lower half-plane is deformed but it still confines the pair cloud. 

\begin{figure}[ht]
\includegraphics[width=\columnwidth]{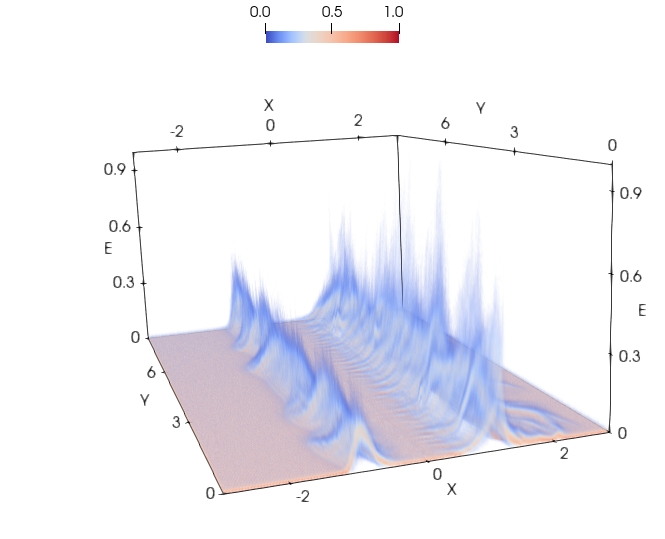}
\caption{Proton phase space distribution at $t=20$. The color scale shows the square root of the density, which is normalized to its peak value. The energy E is expressed in MeV.}
\label{figure7}
\end{figure}

Figure~\ref{figure7} presents the phase space density distribution associated with the proton density distribution in Fig.~\ref{figure6}(d).
We find solitary waves at the positions of the initial EMPs. Protons at the crests of the oscillations reach energies of about 300 keV. This energy corresponds to the speed $v_p=c/40$, which confirms the estimate by Eqn.~\ref{estimate} and underlines that proton inertia sets the speed of an EMP. Some protons are reflected by the EMP that moves to increasing $x$. They are accelerated to $2v_p$, which gives them the energy of 1 MeV in the simulation frame. Figure~\ref{figure2} demonstrates that the proton density peak coincides with that of the magnetic pressure, which is characteristic of fast magnetosonic modes. 

The ion-acoustic speed is $c_s={(k_B(\gamma_eT_e+\gamma_pT_p)/m_p)}^{1/2}$. Ion-acoustic oscillations are slow on electron time scales, which allows electrons to interact with many such waves during one oscillation and be scattered by plasma thermal noise. Ion acoustic waves accelerate protons only in the direction of the wave vector. It is thus widely assumed that electrons have three degrees of freedom and protons one on the time scales of interest. These degrees of freedom give the adiabatic constants $\gamma_e=5/3$ and $\gamma_p=3$, respectively. The electron and proton temperatures in the ambient plasma are both $T_0$ and $c_s\approx 3.2 \times 10^{-3}c$. The Alfv\'en speed $v_A=B_0/{(\mu_0n_0m_p)}^{1/2}$ in the ambient plasma is $v_A \approx 2\times 10^{-3}c$. 

The speed $v_p$ of the EMP is about 6.6 times the fast magnetosonic speed $v_{fms}={(c_s^2+v_A^2)}^{1/2}$. The solitary wave in the proton distribution is thus way too fast to be a fast magnetosonic soliton. It has this speed because it is accelerated continuously by the electric field of the EMP. At early times and near $x=0$, the electric field has a low amplitude and it hardly accelerates protons. In time, the EMP moves away from $x=0$ and its electric field and the proton velocity change increase, which results in the observed proton energy profile close to the boundary at $x=0$ in Fig.~\ref{figure7}.

\section{\label{late}Long term evolution}

It is important to know if the transition layer, which forms after the collapse of the initial EMP, is able to maintain a separation of the pair plasma and the ambient plasma or at least slow down their mixing. If this is the case, the transition layer still acts as a discontinuity. In what follows, we track the evolution of the interfaces between the pair cloud and the ambient plasma. We consider first the interface in the half-space $x>0$.

\subsection{Forward expansion}

\begin{figure*}[ht]
    \includegraphics[width=\textwidth]{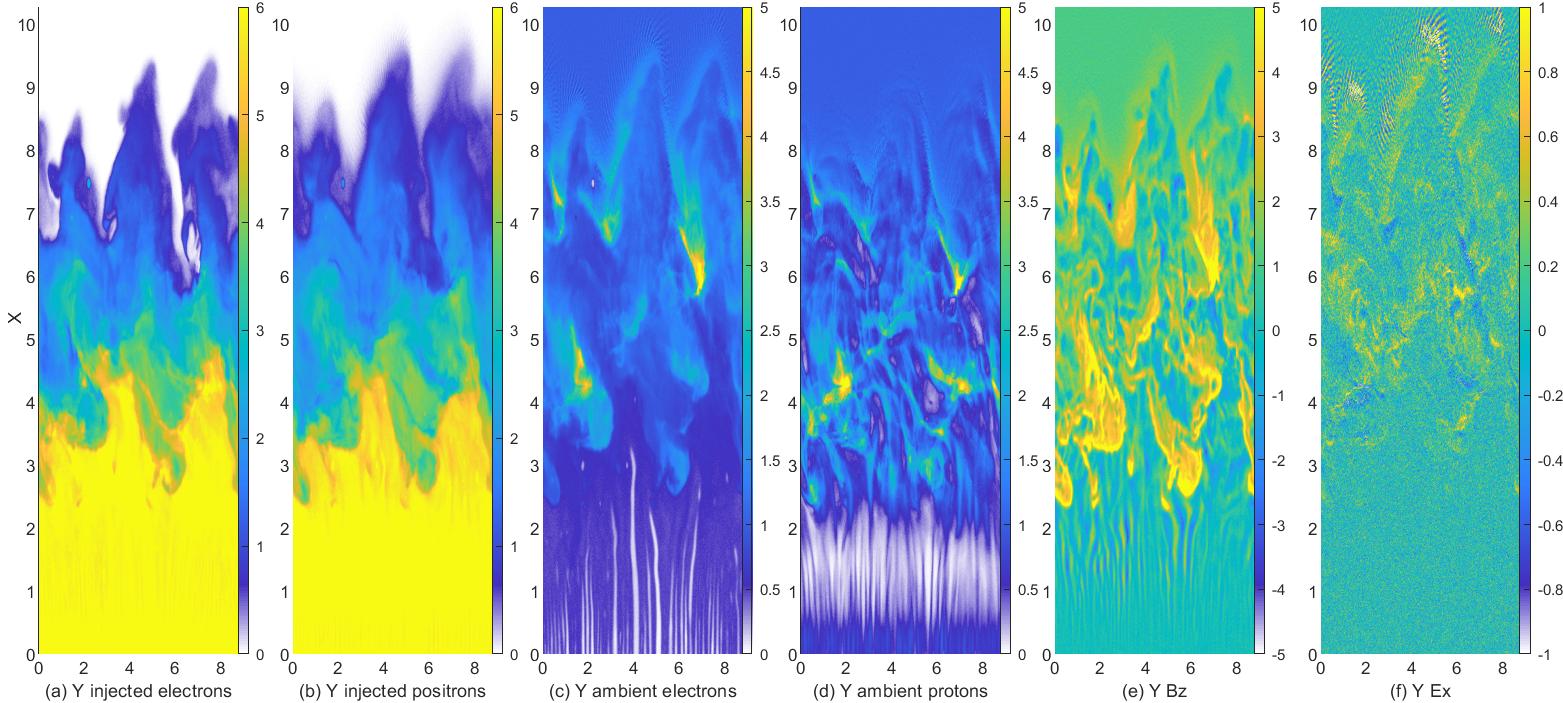}
    \caption{Plasma and field distribution at $t=100$ in the interval $x\ge 0$: Panels (a, b) show the densities of the injected electrons and positrons. Densities of the ambient electrons and protons are displayed in panels (c, d). Panel (e) and (f) show $B_z$ and $E_x$. Densities are normalized to $n_0$, the magnetic field to $B_0$ and the electric field to $cB_0$.}
    \label{figure8}
\end{figure*}

Figure~\ref{figure8} presents the plasma and field data at $t=100$. 
The plasma and field distributions show three domains. Domain 1 with $0 \le x \le 3$ is characterized by a dense pair cloud with a per species density of about $6$ and a low density of the ambient plasma. The amplitude of the initial EMP grew in time and it became strong enough to accelerate and expel protons at $x\approx 0.5$. Hardly any protons are left in the interval $0.5 \le x \le 2$.  Domain 3 is the ambient plasma, which has not yet been affected by the expanding pair cloud. The electromagnetic fields in domains 1 and 3 are not zero. The initial magnetic field is still present in the unperturbed ambient plasma. Statistical fluctuations of the charge density give rise to electric field fluctuations in both outer domains while current density fluctuations are responsible for the magnetic noise in the domain occupied by the pair cloud. Figure~\ref{figure3} demonstrated that the spatial average of the amplitude of these fluctuations is zero in the pair cloud. The fluctuations are strong in the pair cloud and weak in the ambient plasma because the spatially averaged power $B_z^2$  increases with the temperature. Domain 2 is located between the outer two. Pair cloud plasma and ambient plasma coexist in this domain and their interaction drives strong coherent electromagnetic fields and waves.

\begin{figure*}[ht]
    \includegraphics[width=\textwidth]{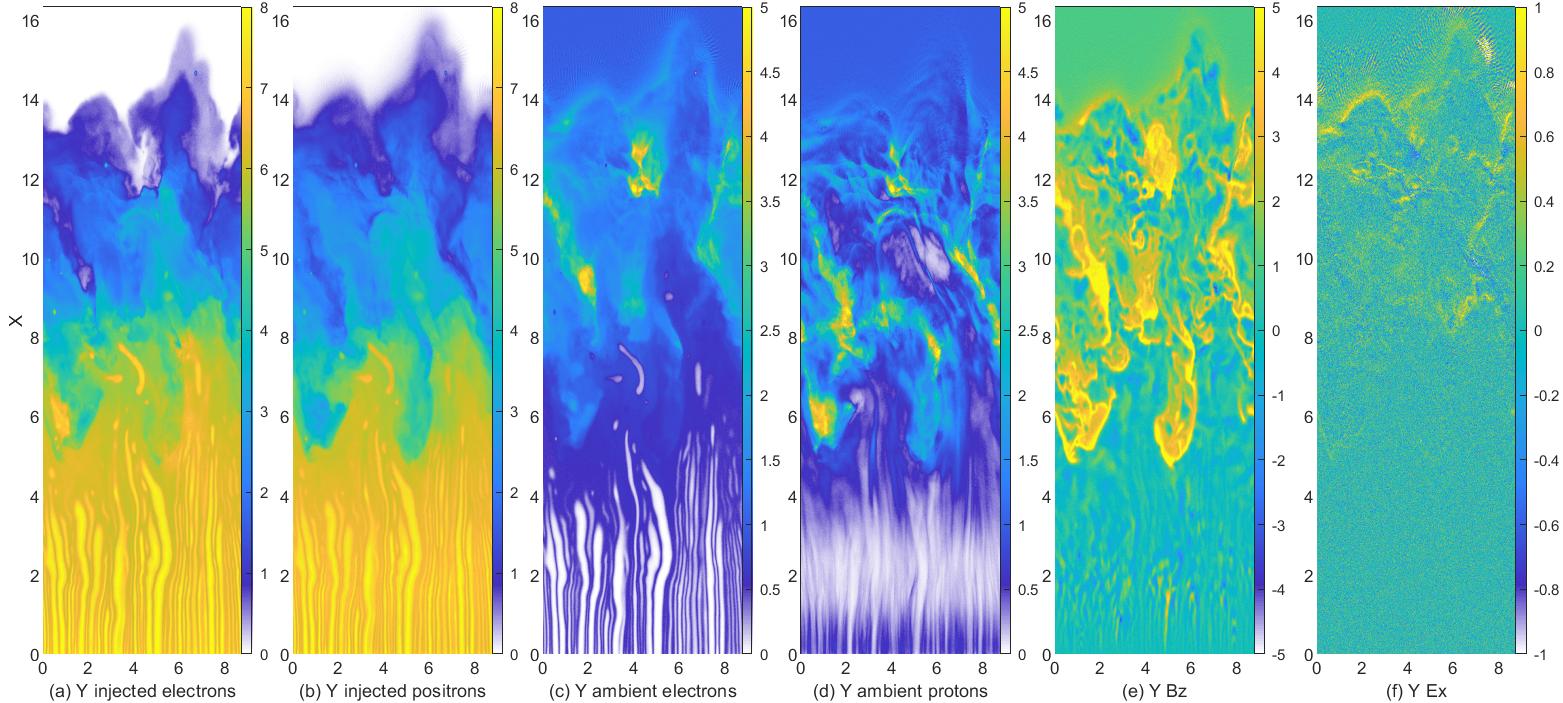}
    \caption{Plasma and field distribution at $T_{sim}=200$ in the interval $x\ge 0$: Panels (a, b) show the densities of the injected electrons and positrons. Densities of the ambient electrons and protons are displayed in panels (c, d). Panel (e) and (f) show $B_z$ and $E_x$. Densities are normalized to $n_0$, the magnetic field to $B_0$ and the electric field to $cB_0$ (Multimedia view).}
    \label{figure9}
\end{figure*}

Figure~\ref{figure9} (Multimedia view) animates the distributions of $B_z$, $E_x$ and the plasma densities in time and shows them at the time $T_{sim}=200$. We find the same subdivision into three domains of the plasma and field distributions. Domain 3, which is ambient plasma that has not yet been affected by the pair plasma, is found at large $x$. Most protons have been expelled from domain 1 in $1 \le x \le 5$. This interval has been filled by a dense pair plasma with a mean positron density of about 6. Fingers, which extend from the boundary $x=0$ into the pair cloud, reach the peak density $8$. Their origin is an instability between the pair cloud and the protons near the boundary. Their large inertia lets protons react slowly to the streaming pair plasma. In time, filaments form in the proton density distribution at  $0 \le x \le 1$ in Fig.~\ref{figure9}(d). Pair cloud particles, which stream across these filaments, must maintain quasi-neutrality. They rearrange themselves into filaments, which emerge in the form of fingers. These fingers are confined by an in-plane magnetic field (not shown). 

Domain 2 is the transition layer between the pure pair plasma and the ambient plasma. It is characterized by a clumpy proton distribution, which is found in the interval $2.5 \le x \le 9.5$ in Fig.~\ref{figure8} and in the interval $5 \le x \le 15$ in Fig.~\ref{figure9}. Its center along $x$ has thus propagated from $x\approx 6$ to $x\approx 10$ during 100 time units, which yields the speed $0.04c$. It exceeds the speed $v_p = c/40$ of the initial EMP but it remains well below the mean speed $v_d=0.6c$ of the pair cloud. The width of the transition layer increases during this time from 7 to 10, which yields the expansion speed $0.03c$. The transition layer slows down the pair cloud's expansion by a factor of 15 and mixing between both species is slow.

The thickness of the transition layer exceeds by far the gyroradius $\approx 0.2$ of leptons moving at the speed $v_d$ in a magnetic field with the strength $B_0$. Figure~\ref{figure9}(e) demonstrates that the pair plasma created channels, from which the coherent background magnetic field was expelled. The pair plasma streams freely through these channels devoid of magnetic fields. Ambient electrons that were pushed forward by the pair plasma created the electric field at the pair cloud's boundary, changing it into an EMP.  

Patches within domain 2, which are filled with a strong magnetic field, coincide with clumps of ambient plasma in Figs.~\ref{figure9}(c-e). Figure~\ref{figure9}~(Multimedia view) reveals that these clumps are ambient plasma, which was displaced and compressed by the expanding fingers of pair plasma. Some patches are what remains from the initial EMP but the expanding pair cloud also creates magnetized clumps of ambient plasma well ahead of the location of the initial EMP. The magnetic field within these patches reaches its peak amplitude on the side that is facing the inflowing pair cloud. Ambient plasma has a residual magnetic field that points in the positive z-direction. It deflects electrons and positrons of the pair cloud into opposite directions and the ensuing net current amplifies the residual field to a peak amplitude that is more than 5 times larger than $B_0$. Figure~\ref{figure8} and Fig.~\ref{figure9} demonstrate that these strong magnetic fields are not driving detectable coherent electric fields and that they cannot fully separate protons from positrons. We note in this context that, unlike the magnetized ambient plasma ahead of the initial EMP, the size of the proton and magnetic field accumulations is not large compared to a gyroradius of a lepton with the speed $v_d$. These magnetic boundaries are thus not EMPs. Given that the magnetic pressure associated with the amplified magnetic fields is an order of magnitude larger than the initial thermal pressure $2P_e$ of the ambient plasma, protons will react to the magnetic pressure gradient force. 

\subsection{Backward expansion}

\begin{figure*}[ht]
    \includegraphics[width=\textwidth]{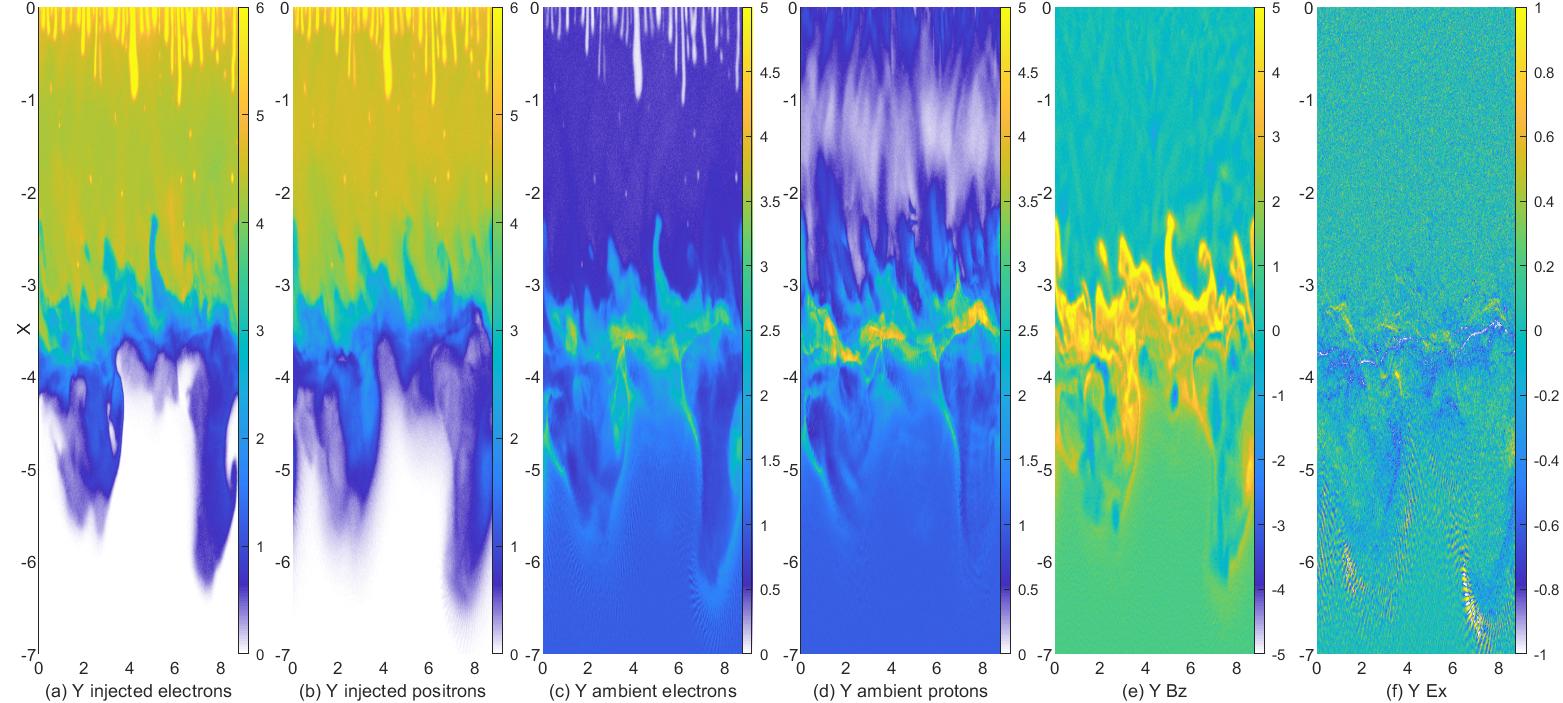}
    \caption{Plasma and field distribution at $t=100$ in the interval $x\le 0$: Panels (a, b) show the densities of the injected electrons and positrons. Densities of the ambient electrons and protons are displayed in panels (c, d). Panel (e) and (f) show $B_z$ and $E_x$. Densities are normalized to $n_0$, the magnetic field to $B_0$ and the electric field to $cB_0$.}
    \label{figure10}
\end{figure*}

Figure~\ref{figure10} presents the field and plasma distribution at the time $t=100$. The front of the dense part of the pair cloud at $x\approx -3$ is approximately planar. Fingers in the pair cloud density emerge at the boundary $x= 0$ and the longest have reached the position $x=-1$. Their cause is the aforementioned instability between the pair cloud and the protons, which were too close to the boundary to be accelerated by the initial EMP. 
The proton density in Fig.~\ref{figure10}(d) peaks at $-4\le x \le -3$. These protons were piled up by the initial EMP. 

Although the initial EMP was also destroyed by the streaming instability, the transition layer that emerged out of it remained more compact than the forward-moving one. A strong magnetic field has developed on the side of the proton accumulation that faces the pair cloud flow. It has been amplified by the current of the injected pair particles that drifted in the residual magnetic field of the ambient plasma. Electric fields with strength $-0.5$ have been induced by ambient electrons, which were pushed downwards by the expanding pair cloud. Only weak electromagnetic fields exist in domain 1 above the initial EMP, which contains the pair cloud and only a small number of protons. Pair plasma streamed through the deformed initial EMP and expanded far upstream of domain 1 in the form of two fingers, the larger of which reached $x\approx -6.5$ at $y\approx 8$. As before, ambient electrons were expelled by the moving magnetic field in Fig.~\ref{figure10}(e) that kept them separated from the injected electrons. Consequently, the density of the ambient electrons is reduced inside both fingers and increased to about 1.5-2 at their boundaries. Protons reacted to the electric field induced by the moving ambient electrons and new EMPs grew near the boundaries of the fingers. Domain~2 in $-7 \le x \le -4$ is again characterized by the simultaneous presence of protons, positrons, and strong electromagnetic fields. Pair cloud particles are confined to the fingers while Fig.~\ref{figure8} showed a less orderly distribution of these particles in the transition layer at the front of the forward-moving pair cloud. Figure~\ref{figure10}(f) reveals rapid electric field oscillations near the boundaries of the pair cloud fingers at $x\approx -6.5$ and $y\approx 7$ and at $x\approx -6$ and $y\approx 1.8$. Their electrostatic nature and short wavelength suggest that they arise from the same streaming instability that destroyed the initial EMP. 

\begin{figure*}[ht]
    \includegraphics[width=\textwidth]{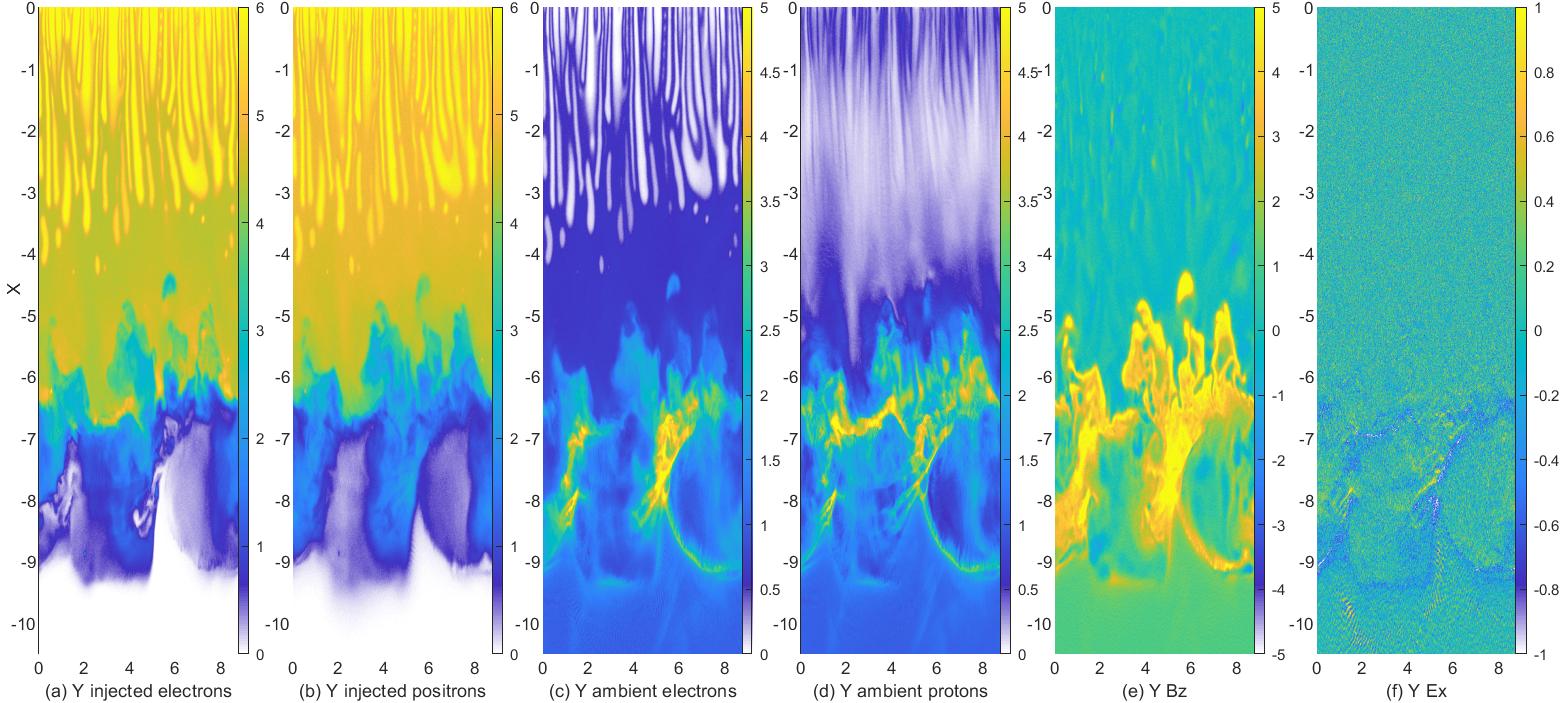}
    \caption{Plasma and field distribution at $T_{sim}=200$ in the interval $x\le 0$: Panels (a, b) show the densities of the injected electrons and positrons. Densities of the ambient electrons and protons are displayed in panels (c, d). Panel (e) and (f) show $B_z$ and $E_x$. Densities are normalized to $n_0$, the magnetic field to $B_0$ and the electric field to $cB_0$ (Multimedia view).}
    \label{figure11}
\end{figure*}

Figure~\ref{figure11} shows the plasma and field distribution at $T_{sim}=200$. The front of the dense part of the pair cloud (domain 1) propagated from $x=$ -3 to -6. Unlike the case shown by Fig.~\ref{figure9}, we observe periodic stable structures in domain 2, which is now located in the interval $-9 \le x \le -6$. These structures can be seen in all displayed plasma and field components.
Figure~\ref{figure11}~(Multimedia view) evidences their stability. The lower pressure exerted by the pair cloud on the ambient plasma in the half-space $x<0$ leads to a less turbulent structure of domain~2 compared to the one in Fig.~\ref{figure9}. We use again the distribution of proton clumps to quantify the mean speed and expansion speed of the transition layer. We find such clumps in the interval $-6 \le x \le -1.5$ at $t=100$ and in the interval $-9 \le x \le -4$ at $t=200$, giving a mean speed modulus $2.75 \times 10^{-2}c$ and expansion speed $1.25 \times 10^{-2}c$. 

\begin{figure}[ht]
\includegraphics[width=\columnwidth]{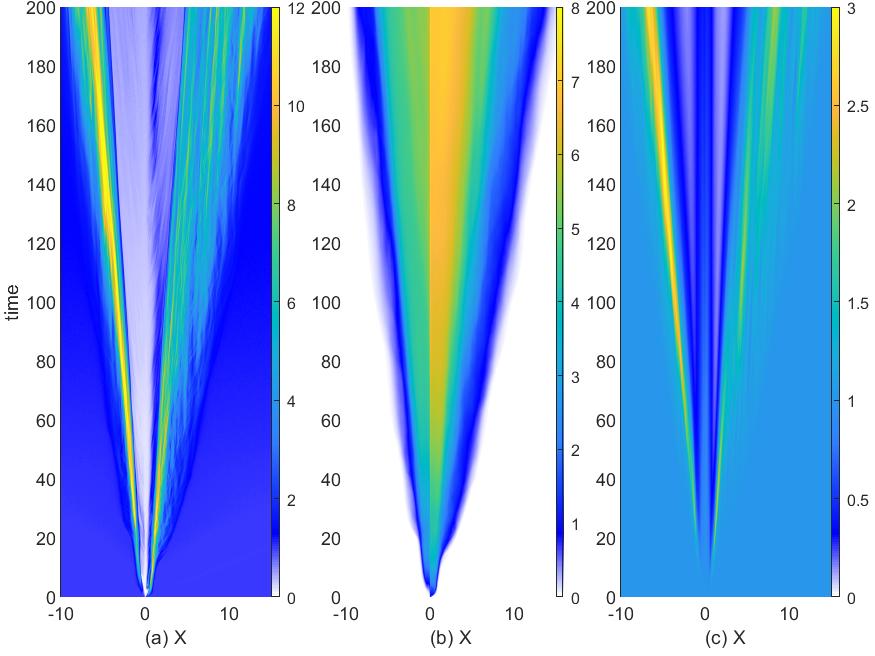}
\caption{Y-averaged quantities: Panel (a) shows $P_{mag}(x,t) = B_z^2(x,t)/2\mu_0P_e$, where $P_e$ is the initial thermal pressure of the ambient electrons. Panels (b) and (c) show the densities of positron and protons, respectively.}
\label{figure12}
\end{figure}

Figure~\ref{figure12} shows how $P_{mag}(x,t) = B_z^2(x,t)/2\mu_0P_e$ and the densities of positrons and protons evolve in time. All quantities were averaged over $y$. 
Before $t\approx 15$, the positron density in Fig.~\ref{figure12}(b) has well-defined fronts on both sides of $x=0$; the pair cloud is confined on both sides by planar initial EMPs. The proton reaction to the expanding pair cloud becomes strong at $t\approx 15$. After $t=15$, we observe qualitatively different interactions between the pair cloud and the ambient plasma on both sides of $x=0$. A diffuse transition layer exists in the half-space $x>0$. It broadens rapidly during $15 \le t \le T_{sim}$ and extends up to $x\approx 15$ at the time $T_{sim}$, which gives its front the mean expansion speed $\approx v_d/8$. Positrons in the interval $x>0$ reached their peak density in those intervals, from which the protons were expelled. Narrow peaks of $P_{mag}$ and proton density exist in the half-space $x<0$. They have reached the position $x\approx -6$ at $t=200$, giving them a propagation speed $\approx v_d/20$. The speeds of the fronts of the forward and backward moving pair clouds correspond to the sums of the mean speeds and expansion speeds of the transition layers, which we estimated from the distributions of the proton clumps. The initial EMP, which propagates to the left at an almost constant speed, is effective at swiping out the protons. Even though the spatially uniform initial magnetic field has been expelled by the expanding pair cloud, we get a value $P_{mag} \approx 0.5$ in both domains~1 around $x=0$ due to the strong incoherent thermal fluctuations.

\subsection{Particle distribution functions}

% change at 126

We determine the energy, to which protons were accelerated during the simulation, and assess how their acceleration affected the energy distributions of the leptons. It is useful to give some reference values for the lepton energy. Leptons that move with mean speed $v_d$ of the pair cloud have an energy of about 130 keV. A pair cloud particle, which moves at the thermal speed $\approx 0.45c$ in the rest frame of the pair cloud, has the speed 0.82 $c$ and energy 750 keV in the simulation frame.  

\begin{figure}[ht]
\includegraphics[width=\columnwidth]{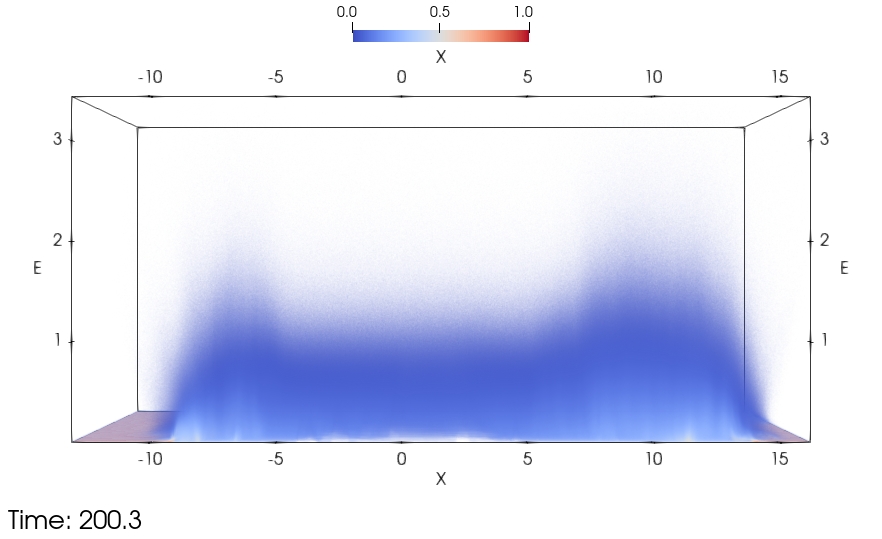}
\caption{Electron phase space density distribution at $T_{sim}=200$. The color shows the square root of the density, which is normalized to its peak density. Energy $E$ is expressed in MeV (Multimedia view).}
\label{figure13}
\end{figure}

\begin{figure}[ht]
\includegraphics[width=\columnwidth]{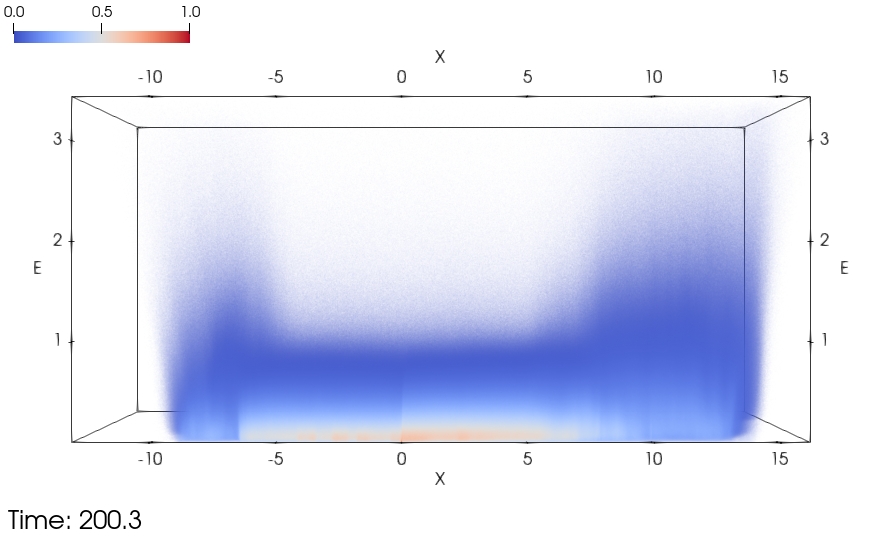}
\caption{Positron phase space density distribution at $T_{sim}=200$. The color shows the square root of the density, which is normalized to the peak density of the electrons. Energy $E$ is expressed in MeV (Multimedia view).}
\label{figure14}
\end{figure}

\begin{figure}[ht]
\includegraphics[width=\columnwidth]{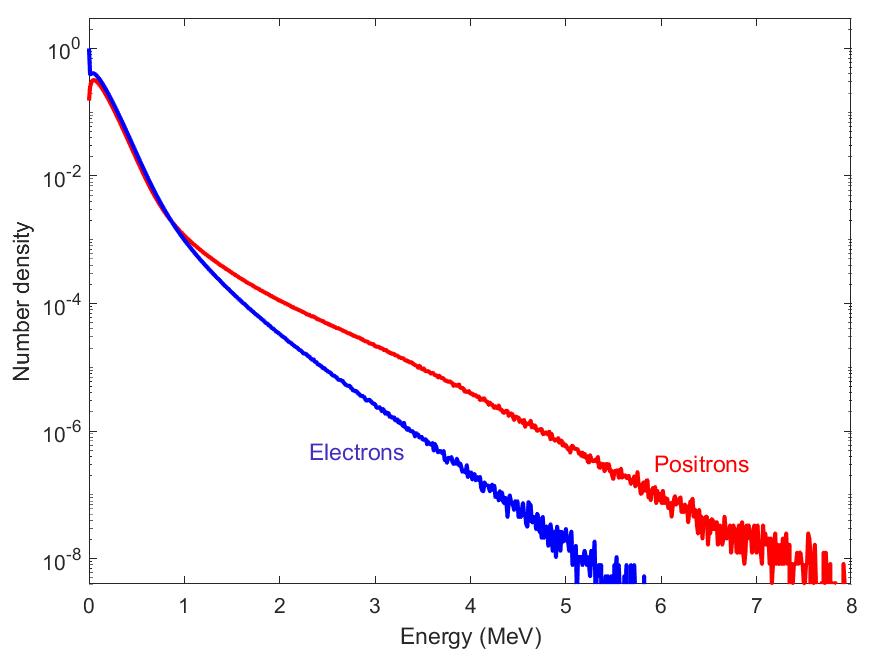}
\caption{Box-integrated energy distributions of electrons and positrons at the time $T_{sim}=200$. Both curves are normalized to the peak value of the electron curve.}
\label{figure15}
\end{figure}

Figure~\ref{figure13}~(Multimedia view) and Fig.~\ref{figure14}~(Multimedia view) follow the energy distributions of electrons and positrons in time. Both lepton species have a diffuse and spatially uniform energy distribution for $-5\le x \le 5$ at $T_{sim}=200$, reaching peak energy of just above 1 MeV. Their energy range is the one expected for the injected pair cloud particles. Electrons and positrons have a similar energy spread. We identified the interval $-5 \le x \le 5$ as domain 1, in which positrons contribute most of the positive charge. The energy spread of electrons and positrons increases in the transition layers near both fronts of the positron cloud. Significant numbers of electrons reach energies of about 2 MeV while positrons reach and exceed the maximum of the displayed energy range. 
We attribute this difference to the electric field of the EMPs near proton accumulations in the transition layer or at its front. This electric field accelerates protons and positrons and decelerates injected electrons in the expansion direction of the pair cloud, which caused also the different density distributions of both pair cloud species in Fig.~\ref{figure5}. 
Figure~\ref{figure15} compares the energy distributions of the electrons and positrons, which were integrated over the box. 
The peak in the electron distribution at low energies does not have a positronic counterpart; these are the ambient electrons outside the transition layer. Both curves follow each other closely between about 100 keV and 1 MeV. These are mostly pair cloud particles that have not interacted yet with the ambient plasma and are close to thermal equilibrium. Both species show an exponential fall-off at large energies with electrons showing a faster decrease. These energetic particles were accelerated in the transition layers. 

\begin{figure}[ht]
\includegraphics[width=\columnwidth]{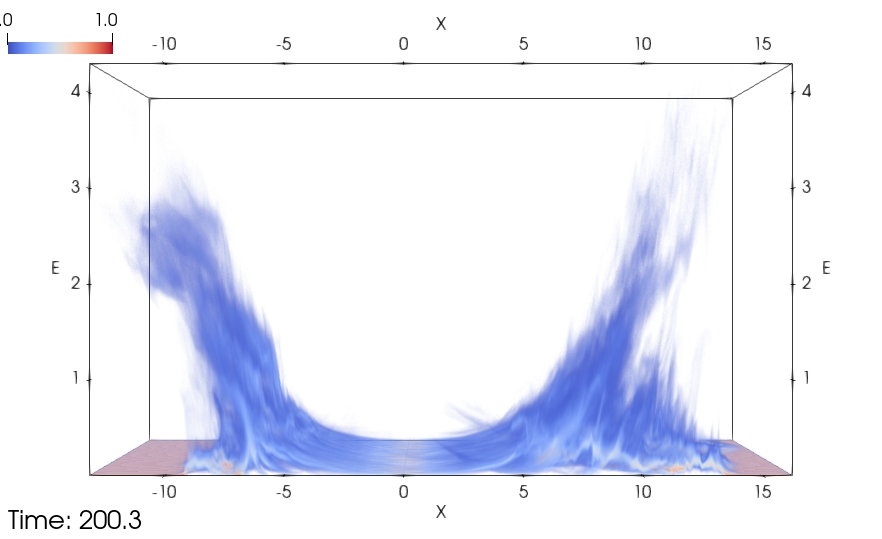}
\caption{Proton phase space density distribution at $T_{sim}=200$. The color shows the square root of the density, which is normalized to its peak density. Energy $E$ is expressed in MeV (Multimedia view).}
\label{figure16}
\end{figure}

The solitary waves, which we observed in Fig.~\ref{figure7} before the initial EMPs were destroyed, do not exist anymore at $T_{sim}=200$ in Fig.~\ref{figure16}. They have been replaced by broad layers, in which protons have been accelerated. Despite their different structure, both transition layers accelerate protons to about the same peak energy of about 4 MeV.

Figure~\ref{figure12} demonstrated that the fronts of the transition layers expanded at the respective speeds $v_d/8$ and $v_d/20$. Protons moving at these speeds have the kinetic energies of 2.6 MeV and 420 keV. The fastest protons in Fig.~\ref{figure16} outrun both transition layers provided that their velocity vector is aligned with the expansion direction of the transition layer. These protons will eventually interact with the ambient plasma either through beam instabilities or by means of a fast magnetosonic shock thereby creating an outer cocoon filled with hot protons. Fast magnetosonic shocks form and evolve over time scales $\omega_{ci}^{-1}\gg T_{sim}$, which we cannot resolve with a 2D simulation. 

Figures~\ref{figure8}-\ref{figure11} revealed a domain structure similar to that of the hydrodynamic jet sketched in Fig.~\ref{figure1}. Domain 1 filled with unmagnetized, dense, and slowly expanding pair plasma corresponds to the inner cocoon. The mean speed of the pair cloud is set by the propagation speed $v_p \ll v_d$ of its front, where pair cloud particles are reflected. The unperturbed ambient plasma in the simulation (domain 3) and in the hydrodynamic jet model are the same; an outer cocoon bounded by an external shock could not form during the simulation time. 

\section{Discussion}

We examined with a 2D simulation the expansion of a pair cloud into a magnetized ambient plasma. The pair plasma expelled the magnetic field and piled it up ahead of it. The piled-up magnetic field trapped ambient electrons and pushed them into the expansion direction of the pair cloud. On the short time scale resolved by the simulation, the protons could not react to the magnetic field. They were thus unable to balance the current of the moving ambient electrons. An electric field was induced that accelerated protons into the expansion direction of the pair cloud. We referred with electromagnetic pulse (EMP) to the combination of a magnetic field, which confines the pair cloud, and the electric field that accelerates the protons. Our periodic boundaries allowed pair plasma to cross it and also expand into the ambient plasma on the other side of the simulation box. This setup allowed us to study with one simulation the expansion into the ambient plasma of two pair clouds and EMPs with different bulk properties. 

Initially, the EMPs at the fronts of both expanding pair plasma clouds were planar due to the uniform injection of pair plasma. The propagating EMPs grew from a small amplitude and their electric fields could not accelerate protons at early times and near the boundary. After a few inverse proton plasma frequencies, the electromagnetic field of both EMPs became strong enough to couple them to the protons. Their reaction limited the speed of the EMPs. Since we continuously injected pair plasma with a mean speed far greater than that of the EMP, the pair plasma density behind each EMP increased. Eventually, a balance was established between the pressure of the pair cloud and the ram pressure the protons exerted on the moving EMP. The EMP and the accelerated protons moved at a few percent of the speed of light. 

The rapid drift of the ambient electrons at the EMP's front and their interaction with the protons resulted in a streaming instability. Electrostatic waves grew in the current sheaths ahead of the EMPs and interacted with the ambient electrons, positrons that had leaked through the EMP, and protons. The interaction of their nonuniform wave fields with the plasma gave rise to a spatially varying dissipation of current density ahead of the initial EMP and, hence, to a spatially varying magnetic field gradient; the EMP could not remain planar. Its deformation grew in time and could not be stabilized by increasing magnetic tension. Such instabilities have also been observed at discontinuities between an expanding plasma and an ambient plasma in space plasmas~\cite{Winske89} and in simulations of laser-generated plasma.~\cite{DieckmannTD}

The following picture emerged. Close to the injection boundary, positrons contributed most of the positive charge. Protons took that role far from the injection boundary. In a jet model, these two domains would correspond to the inner cocoon and the unperturbed ambient plasma. Our simulation was too short to capture the growth of an outer cocoon, which forms on time scales in excess of a few inverse proton gyrofrequencies.~\cite{Dieckmann20} Apart from the magnetic field, which was generated by the instability between the pair plasma and the protons near the injection boundary, the plasma in our inner cocoon was free of any detectable coherent magnetic field. This was expected because we injected unmagnetized pair plasma. Protons and positrons interacted in a transition  layer between both domains. Interactions were mediated by EMPs as well as by strong magnetic fields, which separated the pair plasma from accumulations of ambient plasma. Their origin was a residual magnetic field, which was amplified  by the drift current of the pair cloud. The width of the transition layer was of the order of a few proton skin depths. 

How does this width compare to the size of relativistic jets? Let us assume that the relativistic jet moves through a stellar wind like the one emitted by our sun. At the Earth's orbit, the solar wind density is about 5 $\mathrm{cm}^{-3}$. The particle's mean free path, which sets the thickness of a contact discontinuity, is about 1 astronomical unit.~\cite{Goldstein05} The simulation time $T_{sim}=200$ corresponds to 68 milliseconds and the spatial unit to 100 km. Albeit the thickness of the transition layer $\sim 10$ is large in our simulation, it is 6 orders of magnitude less than the mean free path of the solar wind and, hence, well capable of forming a discontinuity that is infinitesimally thin on jet scales.

Two important questions could not be addressed by our simulation and must be left to future work. Firstly, the transition layer should be susceptible to a magnetized Rayleigh-Taylor instability because a pair plasma is pushing much heavier protons. This instability can grow even if ions are unmagnetized~\cite{Winske96} and if the thickness of the transition layer is comparable to the wavelength of its growing modes.~\cite{Brown88,Hillier16} We do not observe a Rayleigh-Taylor instability here, which may be a result of the spatially nonuniform plasma distribution within the transition layer or the steady increase of its thickness. Secondly, given that the long-term evolution of EMPs is different if the magnetic field in the ambient plasma is oriented in or orthogonal to the simulation plane and the interplay of these modes,~\cite{Hillier16} it would be important to study the structure of the transition layer in a 3D geometry. 

\section*{CONFLICT OF INTEREST}

The authors have no conflicts to disclose.

\section*{ACKNOWLEDGEMENTS}
The simulations were performed on resources provided by the Swedish National Infrastructure for Computing (SNIC) at the NSC and on the centers of the Grand Equipement National de Calcul Intensif (GENCI) under grant number A0090406960. The first author also acknowledges financial support from a visiting fellowship of the Centre de Recherche Astrophysique de Lyon.

\section*{DATA AVAILABILITY}
The data that support the findings of this study are available from the corresponding author upon reasonable request.

\section*{REFERENCES}

%\nocite{*}
\bibliography{manuscript}% Produces the bibliography via BibTeX.

\end{document}